# Computer-free, all-optical reconstruction of holograms using diffractive networks

Md Sadman Sakib Rahman[1,2,3] and Aydogan Ozcan[1,2,3,*]

[1]Electrical and Computer Engineering Department, University of California, Los Angeles, CA, 90095, USA

[2]Bioengineering Department, University of California, Los Angeles, CA, 90095, USA

[3]California NanoSystems Institute (CNSI), University of California, Los Angeles, CA, 90095, USA

[*]Corresponding author: ozcan@ucla.edu

**Abstract**

Reconstruction of in-line holograms of unknown objects in general suffers from twin-image artifacts due to the appearance of an out-of-focus image overlapping with the desired image to be reconstructed. Computer-based iterative phase retrieval algorithms and learning-based methods have been used for the suppression of such image artifacts in digital holography. Here we report an all-optical hologram reconstruction method that can instantly retrieve the image of an unknown object from its in-line hologram and eliminate twin-image artifacts without using a digital processor or a computer. Multiple transmissive diffractive layers are trained using deep learning so that the diffracted light from an arbitrary input hologram is processed all-optically, through light-matter interaction, to reconstruct the image of an unknown object at the speed of light propagation and without the need for any external power. This passive all-optical processor composed of spatially-engineered transmissive layers forms a diffractive network, which successfully generalizes to reconstruct in-line holograms of unknown, new objects and exhibits improved diffraction efficiency as well as extended depth-of-field at the hologram recording distance. This all-optical hologram processor and the underlying design framework can find numerous applications in coherent imaging and holographic display-related applications owing to its major advantages in terms of image reconstruction speed and computer-free operation.

**Introduction**

Holography is a widely used technique with a myriad of applications in e.g., computational imaging, displays, interferometry and data storage [1]. What distinguishes holography from other optical methods is its ability to record and reconstruct both the intensity and the phase of the object field of interest. For its recording, the object wave is made to interfere with a reference wave to generate an intensity pattern, i.e., the hologram, encoding both the amplitude and the phase information of the object wave. Holographic reconstruction broadly refers to the retrieval of the object information from the recorded hologram intensity.

In the original hologram recording scheme demonstrated by Gabor, the so-called 'in-line holography', the reference wave and the object wave co-propagate along the same axis [2]. In its analog implementation, the recorded hologram (e.g., a photographic plate or its digital copy) can be illuminated with a reference wave and the object field can be partially recovered. Reconstruction of in-line holograms in general suffers from a spatial artifact known as the twin-image, which is due to the appearance of an out-of-focus image overlapping with the desired image of the object, degrading the reconstruction quality. To get rid of the twin-image problem, Leith and Upatniek proposed an alternative hologram recording geometry, known as the 'off-axis holography' [3]. In this scheme, a small angle is introduced between the reference wave and the object wave, resulting in a spatial separation of the twin-image from the desired image of the unknown object during the reconstruction process. However, in addition to the relatively increased complexity of the experimental setup, the achievable space-bandwidth product with off-axis holography is smaller. In fact, the simplicity and the experimental robustness of the recording geometry of in-line holograms make it ideal for various field-based measurements and applications that require compact and cost-effective imaging and sensing systems [4].

Powered by modern computers, the emergence of digital holography enabled the numerical reconstruction of holograms, exploiting the availability of phase retrieval algorithms for suppressing twin-image artifacts [5]. Various phase-retrieval algorithms have been reported over the last several decades to recover the missing phase information and reconstruct an image of the specimen from one or more in-line holograms using a computer [6–12]. As an alternative approach, deep learning-based hologram reconstruction methods have also been demonstrated, performing phase recovery and twin-image elimination using trained deep neural networks [13–22]. Some of these earlier neural networks devised to blindly reconstruct an arbitrary hologram were trained with examples of holograms (input to the network) and the corresponding object fields (target ground truth) [13]. Although the data generation and the training process is time consuming (taking e.g., 12-24 h, depending on the availability of graphics processing units, GPUs), this is still a one-time effort, and once the neural network has been trained, it can be deployed to blindly reconstruct an input hologram of an unknown, new object, in a single feed-forward through the network, without any iterative phase retrieval or optimization steps. This constitutes one of the important advantages of deep learning-based hologram reconstruction methods, in addition to enabling some other unique imaging capabilities such as e.g., extended depth-of-field [23] and virtual image transformations [24,25].

All of this earlier body of work is based on digital processing of the acquired holograms through a computer in order to reconstruct the images of unknown objects. In this work, we present an all-optical hologram reconstruction method that processes the diffracted wave from a holographic recording through a series of transmissive diffractive layers/surfaces that collectively project an image of the unknown object at an output plane. These diffractive layers are designed through deep learning to specifically reconstruct in-line holograms, and form a passive optical network [26–32] that can execute a desired task between an input and output plane without an external power source, except for the illumination light at a wavelength of λ. The input optical field of a diffractive network is transformed via light-matter interactions and diffraction through the spatially-engineered layers to produce the target field at its output plane, dictated by the inference task that is statistically learned. In addition to computing the desired output field at the speed of light propagation (between the input and output planes), this diffractive network architecture also exploits the large connectivity and parallelism of free-space optical diffraction and the advantages of layer-by-layer optical processing [30].

By training five successive diffractive layers using deep learning, we report the design of an all-optical processor, in the form of a passive diffractive network extending only $\sim 225\lambda$ in the axial direction, that can reconstruct the image of an unknown object from its intensity hologram without the need for any digital computation or an external power source. We show that the designed diffractive networks can generalize very well to unseen examples and accurately reconstruct their images at the speed of light propagation, also exhibiting an enhanced diffraction efficiency as well as an improved depth-of-field at the hologram recording distance. We believe that this all-optical hologram processor will find various applications in holographic imaging and display related applications, especially benefiting from its computer-free and instantaneous image reconstruction capability.

**Results**

A schematic of our all-optical hologram reconstruction framework is shown in Fig. 1(a). The hologram recording process is the same as for in-line holography, i.e., an unknown object is illuminated with a coherent plane wave of wavelength $\lambda$, and the recording medium, e.g., an opto-electronic image sensor-array or a photographic emulsion, is placed at a distance $z_0$ from the object plane; in this work, we used $z_0 = 30\lambda$. The component of the wave scattered by the object interferes with the unscattered, directly transmitted wave, producing an in-line hologram. During the all-optical reconstruction through a diffractive network (Fig. 1(b)), the hologram recording medium or its digital copy is assumed to be illuminated with a plane wave of the same wavelength. In this embodiment, a set of spatially-engineered, transmissive diffractive layers is placed between the input hologram and the output (i.e., the reconstruction or observation) plane, which all-optically computes a twin-image-free reconstruction of the original object at the network output (see Fig. 1(b)). The axial distances between successive transverse planes i.e., the input hologram plane, diffractive network layers and the output/reconstruction plane, are denoted by $z_i$, $i = 1,\ldots,N+1$, where $N$ is the number of layers in the diffractive network; in this work, we used $N = 5$ and $z_i = 37.5\lambda$. The size of the unknown object to be reconstructed is assumed to be $25\lambda$, the lateral extent of the recorded hologram is $42\lambda$, and the width of each diffractive layer is assumed to be $100\lambda$.

Based on these hyper-parameters, we trained an all-optical network with *N*=5 diffractive layers, where each layer had a total of 200 × 200 trainable parameters. Each one of these parameters corresponds to the phase value of a diffractive feature (neuron) over an area of $\lambda/2 \times \lambda/2$. Therefore, using deep learning we optimized a total of 0.2 million independent phase values, spread across *N*=5 transmissive diffractive layers forming an optical network (Fig. 1(b)). For our training image data, we used the MNIST handwritten digit dataset, which was augmented with an additional custom-built image dataset of generic shapes, e.g., patches, circles, straight-line and circular-line gratings etc. (see Supplementary Fig. S1 for examples of these additional training images). The total number of images in the training dataset was 110,000 (55,000 from the MNIST training set, 55,000 from the custom prepared image dataset); none of these images in the training dataset were used in the blind testing of the final diffractive network models. Other details of the network training process, including the loss function, optimization algorithm and training times, are reported in the Methods section.

Once the diffractive network was trained for all-optical reconstruction of input holograms, it was blindly tested using in-line holograms of new, unknown objects. Fig. 2 reports numerical examples of input holograms and the corresponding all-optical reconstructions obtained at the output of the diffractive network, which reveal that twin-image related artifacts that normally appear in free-space propagation of the input hologram are eliminated at the observation/output plane of the diffractive network, as desired. Also note that the images of the letters in the English alphabet were not part of our training image dataset, and therefore the success in the all-optical reconstruction of "UCLA" text from its hologram (see Fig. 2) indicates that the trained diffractive network is capable of generalizing to distributions different from the distribution of the objects used during the training phase.

Next, we explored through numerical simulations the capability of the same diffractive network to all-optically reconstruct and resolve closely spaced objects, composed of two parallel lines with gradually increasing separation; see Fig. 3. For line separations that are equal to or larger than $\lambda$, the diffractive network succeeded in reconstructing the in-line holograms of these test objects and faithfully resolved the two lines from each other as shown in Fig. 3. Also notice that for line separations that are larger than $3.5\lambda$, the free-space propagation of the input holograms presents stronger spurious/artificial drops around the center of the lines due to the strong twin-image artifacts; this causes the appearance of artificial lines that are comparable to the signal strength of the true object lines. Such spatial artifacts due to twin images have been successfully suppressed at the output plane of the diffractive network as shown in Fig. 3.

In addition to all-optical reconstruction of in-line holograms and the removal of twin-image artifacts, the presented diffractive network-based holographic image reconstruction framework can provide further advantages such as being more robust to uncontrolled changes in $z_0$, covering an extended depth-of-field in the hologram recording distance. To demonstrate this, we numerically explored the performance of all-optical reconstructions achieved by the same diffractive network design shown in Fig. 1(b) when the hologram recording distance $z_0 = 30\lambda$ was changed from its nominal value, covering an axial range of $z_0 = 24\lambda : 36\lambda$; see Fig. 4. In this case, however, the diffractive network did not know the exact hologram recording distance that was used and therefore its design did not change compared to earlier results so that we could truly test the robustness of its all-optical reconstructions against changes in the hologram recording distance. To quantify the all-optical image reconstruction performance under this change, we used 10,000 test images of the MNIST dataset to calculate the structural similarity index measure (SSIM) and the peak signal-to-noise ratio (PSNR) values for the reconstructed images at the diffractive network's output as shown in Fig. 4(a). Fig. 4(b) also presents numerical simulations for visual comparison of the resulting all-optical reconstructions for different hologram recording distances, revealing the robustness of the diffractive network's reconstructions even though the hologram recording distance was significantly changed (±20%) from its nominal value which was used during the training of the diffractive network.

In fact, the all-optical hologram reconstruction performance of the diffractive network design can be further improved by incorporating the extended depth-of-field in hologram recording distance during its training process so that the hologram reconstruction at the output plane becomes even less sensitive to the exact value of the recording distance. To demonstrate this capability, we trained another diffractive network from scratch, where the hologram recording distance during the training process was treated as a random variable, $z_{train} \sim Uniform\left(\left(1-\delta_{tr}\right)z_0, \left(1+\delta_{tr}\right)z_0\right)$, where $\delta_{tr} = 0.2$ was used (see the Methods section for details). Fig. 4(a) reveals that the resulting new diffractive network design based on this training strategy achieves all-optical reconstruction of input holograms with relatively flat SSIM and PSNR curves, demonstrating its inference success over a large range of hologram recording distances. Fig. 4(b) also supports the same conclusion by reporting the reconstructed holographic images at the output plane of the diffractive network for $\delta_{tr} = 0.2$.

A practical consideration for the use of this presented all-optical hologram reconstruction approach would be the diffraction efficiency of the optical design. The average diffraction efficiencies of the diffractive networks trained with $\delta_{tr} = 0$ and $\delta_{tr} = 0.2$ are found to be ~0.31% and ~0.58% respectively, whereas the average diffraction efficiency for free-space propagation-based reconstruction is ~8.61%. Here, the diffraction efficiency is defined as the ratio of the power at the output reconstruction field-of-view to the input power illuminating the hologram area; the above reported average diffraction efficiency values were calculated over 10,000 test images of the MNIST dataset. In Fig. 5, we show that the diffraction efficiency of the hologram reconstruction diffractive network can be significantly improved by adding an efficiency-related penalty term in the training loss function (see Methods section). By properly adjusting the relative weight ($\eta$) of this additional loss term to 0.5, a diffractive network with an output diffraction efficiency of ~23.64% was designed; its all-optical reconstruction performance is reported in Fig. 5(a). Furthermore, Fig. 5(b) shows the diffraction efficiency-SSIM trade-off that can be achieved by changing $\eta$. For example, by further increasing $\eta$ to 3.0, a hologram reconstruction diffractive network with an average diffraction efficiency of ~26.2% can be designed with a minor sacrifice in the image reconstruction quality.

**Discussion**

Unlike traditional phase recovery and twin image elimination methods that are based on algorithms implemented in digital computers, this work introduces a passive all-optical processor, composed of deep learning-designed diffractive layers, that can collectively reconstruct a hologram at its output plane without any digital computation or an external power source except for the illumination light. In addition to its passive nature, the hologram reconstruction process occurs almost instantaneously, i.e., at the speed of light propagation through a compact diffractive network that extends only $\sim 225\lambda$ in the axial direction (from the input plane to the output).

Despite its major advantages, there are also some limitations of the presented framework. If the hologram recording distance changes drastically compared to the training $z_0$ value, we might need to fabricate a new set of diffractive layers trained by using the new hologram recording distance. However, our results in Fig. 4 confirmed that the all-optical reconstruction quality of a diffractive network remains very good over a relatively broad range of hologram recording distances, even though the network was trained with a fixed $z_0$ value; in fact, this depth range can be further improved by increasing $\delta_{tr}$ in the training process. Another limitation of the presented method is the required 3D alignment and fabrication precision of the diffractive layers. Previous work [33,34] on diffractive networks for all-optical image classification related tasks revealed that one can vaccinate the diffractive network design and make it more robust in its statistical inference accuracy by incorporating potential misalignments, fabrication imperfections and input object distortions as part of the optical forward model as random variables, similar to the strategy that we used earlier for extending the range of hologram recording distances. Furthermore, the performance of diffractive network-based hologram reconstructions with respect to other practical parameters such as e.g., output diffraction efficiency can be fine-tuned and optimized by adjusting the training loss function as illustrated in Fig. 5, presenting a powerful design flexibility.

In conclusion, we introduced a framework for a passive all-optical processor designed using deep learning, that is capable of instantaneously reconstructing artifact-free images of unknown objects from their in-line holograms, without any digital computation. This all-optical computing framework that is composed of successive transmissive diffractive layers generalizes very well to unseen objects, and exhibits image quality, depth-of-field and diffraction efficiency advantages compared to free-space propagation of a recorded hologram. This all-optical processor and the underlying deep learning-based design framework can find numerous applications in holographic imaging and display applications due to its reconstruction speed (driven by the propagation of light) and passive nature (operating without the need for a digital computer or an external power source, except for the illumination light).

**Methods**

**Numerical Model of Diffractive Networks.** In developing the mathematical model of a diffractive network, we index the transverse planes of the system by the letter $l$. For a diffractive network with $N$ diffractive layers between the input and the output planes, $l = 0, 1, 2, \ldots, N, N+1$, where $l = 0$ represents the input hologram plane and $l = N+1$ represents the output (image reconstruction) plane. Furthermore, we denote the complex amplitude of the optical wave before and after the optical modulation imposed by element/pixel $m$ of a diffractive layer $l$ by $u_m^l$ and $v_m^l$, respectively. To clarify, we use $u$ to denote the field before modulation and $v$ to denote the field after modulation by a diffractive layer, while the index in the superscript is used to represent the corresponding diffractive layer and the subscript represents the individual diffractive features of that layer. Then, following the Rayleigh-Sommerfeld formulation of diffraction [35], we can write:

$$u_k^l = \sum_m \frac{z_l}{\left(r_{mk}^l\right)^2} \left( \frac{1}{2\pi r_{mk}^l} + \frac{1}{j\lambda} \right) \exp\left( j \frac{2\pi r_{mk}^l}{\lambda} \right) v_m^{l-1}$$

$$r_{mk}^l = \sqrt{(x_k - x_m)^2 + (y_k - y_m)^2 + z_l^2}$$

Here $\lambda$ is the wavelength of the optical wave; $z_l$ is the axial distance between layers $l-1$ and $l$; $(x_m, y_m)$ and $(x_k, y_k)$ are the transverse coordinates of feature $m$ of layer $l-1$ and feature $k$ of layer $l$, respectively; $v^0$ and $u^{L+1}$ are the optical fields at the input and the output fields-of-view of the diffractive network, respectively.

If the complex-valued transmittance of pixel $k$ of layer $l$ is denoted with $t_k^l$, then assuming the diffractive layers to be thin, we can write $v_k^l = u_k^l \cdot t_k^l$, which can be rewritten as:

$$v_k^l = \left( \sum_m w_{km}^l v_m^{l-1} \right) \cdot t_k^l \tag{1}$$

$$w_{km}^{l} = \frac{z_l}{\left(r_{mk}^{l}\right)^2}\left(\frac{1}{2\pi r_{mk}^{l}} + \frac{1}{j\lambda}\right)\exp\left(j\frac{2\pi r_{mk}^{l}}{\lambda}\right).$$

Eq. (1) resembles a feedforward neural network, where $w_{km}^{l}$ and $t_{k}^{l}$ are the analogues of weights and (multiplicative) biases, respectively. Note that we use the terms diffractive feature and neuron interchangeably in the context of diffractive networks. In contrast with digital feedforward neural networks, however, the weights $w_{km}^{l}$ of diffractive networks, which are constrained by free-space diffraction between the diffractive surfaces, are not trainable. Nonetheless, the multiplicative biases $t_{k}^{l}$ of the neurons are trainable since they are determined by the physical parameters of the diffractive surfaces. For example, for thin diffractive layers, the phase $\varphi_{k}^{l}$ of the complex transmittance $t_{k}^{l}$ of a diffractive layer neuron is related to its height $h_{k}^{l}$ by the equation $\varphi_{k}^{l} = \frac{2\pi(n-1)h_{k}^{l}}{\lambda}$, where $n$ is the refractive index of the diffractive layer material (which is surrounded by air).

Given a training dataset of input-target pairs $(x, y)$ specific to a desired task, the input $x$ can be mapped to the diffractive network input optical field $v^0$, whereas the diffractive network output optical field $u^{L+1}$ is mapped to $\hat{y}$, an estimate of the target $y$. The amplitude and/or the phase of the complex amplitude transmittance values $t_{k}^{l}$ can be optimized by minimizing an error function between the target and the estimate, i.e., $L(y, \hat{y})$, which is achieved using standard deep learning techniques such as stochastic gradient descent and error backpropagation.

**Network architecture.** The hyperparameters related to the diffractive network architecture were set empirically. All the reported diffractive networks have 5 phase-only modulation/diffractive layers, and the transverse dimension of the layers are $100\lambda \times 100\lambda$. The pixel/neuron size is set as $0.5\lambda \times 0.5\lambda$, and the number of trainable parameters (phase values) in each diffractive network was 5×200×200 = 0.2 million. The axial/longitudinal distance between the layers is set as $37.5\lambda$. The input and output planes of the diffractive network were also assumed to be $37.5\lambda$ away from the first and the last layers, respectively.

In computing the holograms, the object aperture (transverse dimension) was assumed to be $25\lambda \times 25\lambda$ and the recording plane was assumed to be $30\lambda$ away from the object plane. The lateral extent of the recorded holograms was limited to $42\lambda \times 42\lambda$. All the simulations for this work were done at $\lambda = 600$ nm. The results, however, are invariant to wavelength given that all the dimensions are scaled proportional to $\lambda$.

**Training.** For training diffractive networks to all-optically perform holographic reconstruction, we computed the in-line holograms corresponding to the training objects by numerically simulating the propagation of a plane wave scattered by the object from the object plane to the hologram recording plane. Assuming that the hologram amplitude transmittance is proportional to the recorded intensity, the input optical field $v^0$ was set to the normalized hologram intensity. The reconstructed object pattern was defined as the intensity of the output optical field $u^{L+1}$, i.e., $\hat{y} = |u^{L+1}|^2$. We defined the training loss function (to be minimized) as:

$$L = L_{pixel} + 1000 L_{fourier} + \eta L_{efficiency}$$

$$L_{pixel} = \frac{1}{N_p} \sum_{p=1}^{N_p} |y_p - \hat{y}_p|$$

$$L_{fourier} = \frac{1}{N_p} \sum_{p=1}^{N_p} |F\{y\}_p - F\{\hat{y}\}_p|^2$$

$$L_{efficiency} = 1 - \frac{P_{out}}{P_{illum}}$$

Here, $y$ refers to the target (object image), and $F\{\cdot\}$ denotes the Fourier transform operation, $P_{illum}$ is the power of the wave illuminating the hologram and $P_{out}$ is the power over the region of interest (i.e., the reconstruction field-of-view) at the output plane. The coefficient $\eta$ of the diffraction efficiency-related loss term ($L_{efficiency}$) is a training hyperparameter which can be used to control the diffraction efficiency of the resulting optical network (see Fig. 5). We used Adam optimization algorithm for minimizing the loss function [36].

For designing the extended depth-of-field diffractive networks, we treated the distance between the object and the hologram recording plane during the training phase as a random variable $z_{train}$, drawn from a uniform distribution centered around the nominal recording distance $z_0$, i.e., $z_{train} \sim Uniform((1-\delta_{tr})z_0, (1+\delta_{tr})z_0)$ where $\delta_{tr}$ quantifies the width of the distribution. During blind testing phase, we defined $\delta z = z_{test} - z_0$ to be the deviation of the recording distance from the nominal distance.

For training the diffractive models, we used the MNIST handwritten digit dataset, augmented with an additional custom-built dataset of generic shapes, e.g., patches, circles, straight-line and circular-line gratings etc. A few examples of the images from this custom-built dataset can be found in Supplementary Fig. S1. The number of images in the training dataset was 110,000 (55,000 from the MNIST training set and 55,000 from the custom prepared image dataset). None of the images used for evaluating the model performances reported in this work were in the training dataset.

The model was implemented and trained using TensorFlow [37]. The Rayleigh-Sommerfeld diffraction integral was computed using the Angular Spectrum method [35]. We used the native TensorFlow implementation of Adam optimization algorithm with the default hyperparameter values, i.e., 0.001 for the learning rate, and 0.9 and 0.999 for the exponential decay rates for the first and the second moments of the gradients, respectively. The models were trained for 50 epochs with a mini-batch size of 4. The training of a model typically took ~8 hours on a GTX 1080 Ti graphics processing unit (GPU, Nvidia Inc.) on a machine running Windows 10 operating system (Microsoft Inc.).

system for large-scale machine learning," in *Proceedings of the 12th USENIX Conference on Operating Systems Design and Implementation*, OSDI'16 (USENIX Association, 2016), pp. 265–283.

**Figures**

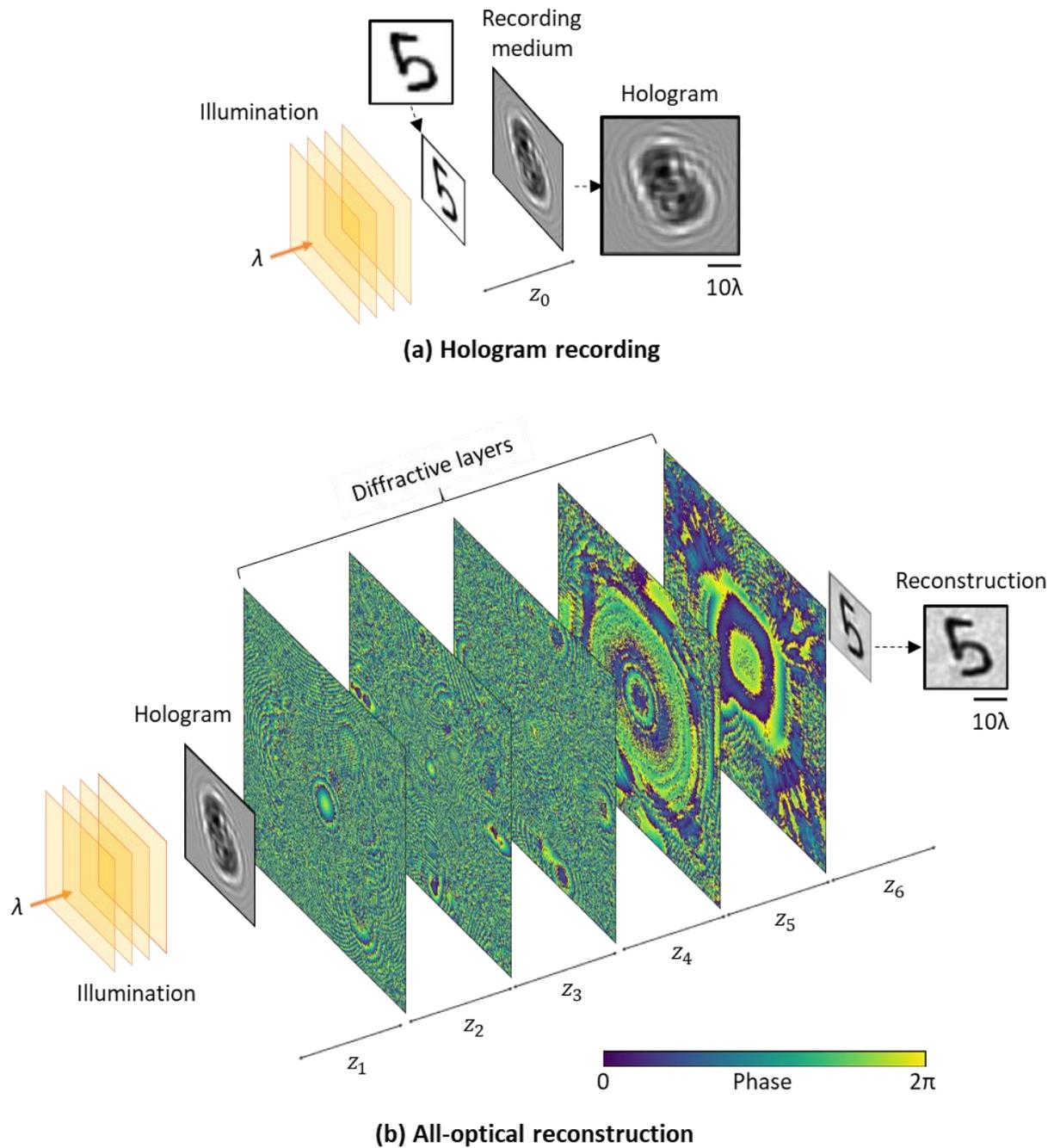

**Fig. 1 Hologram recording and reconstruction scheme for the proposed all-optical holographic reconstruction method.** (a) For hologram recording, the object is illuminated by a plane wave, and the resulting diffraction pattern is recorded, forming the hologram. (b) For reconstruction, the hologram is illuminated with a plane wave, and the diffractive network all-optically computes a twin-image-free reconstruction at the output plane.

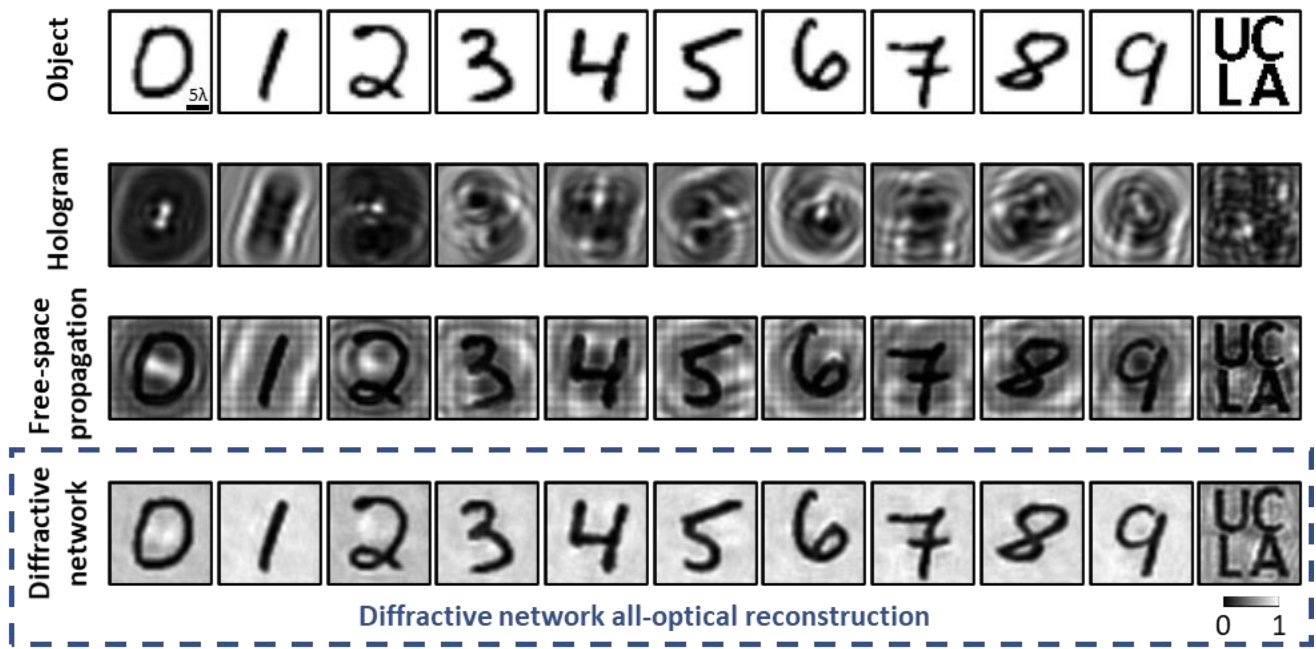

**Fig. 2 Performance of all-optical hologram reconstruction using diffractive networks.** The first two rows depict the target object amplitudes, and the corresponding recorded in-line holograms, respectively. The axial distance between the object and the recording plane is assumed to be 30λ. The reconstructed object intensities by free-space propagation are shown in row 3, whereas those by the designed diffractive network are shown in row 4.

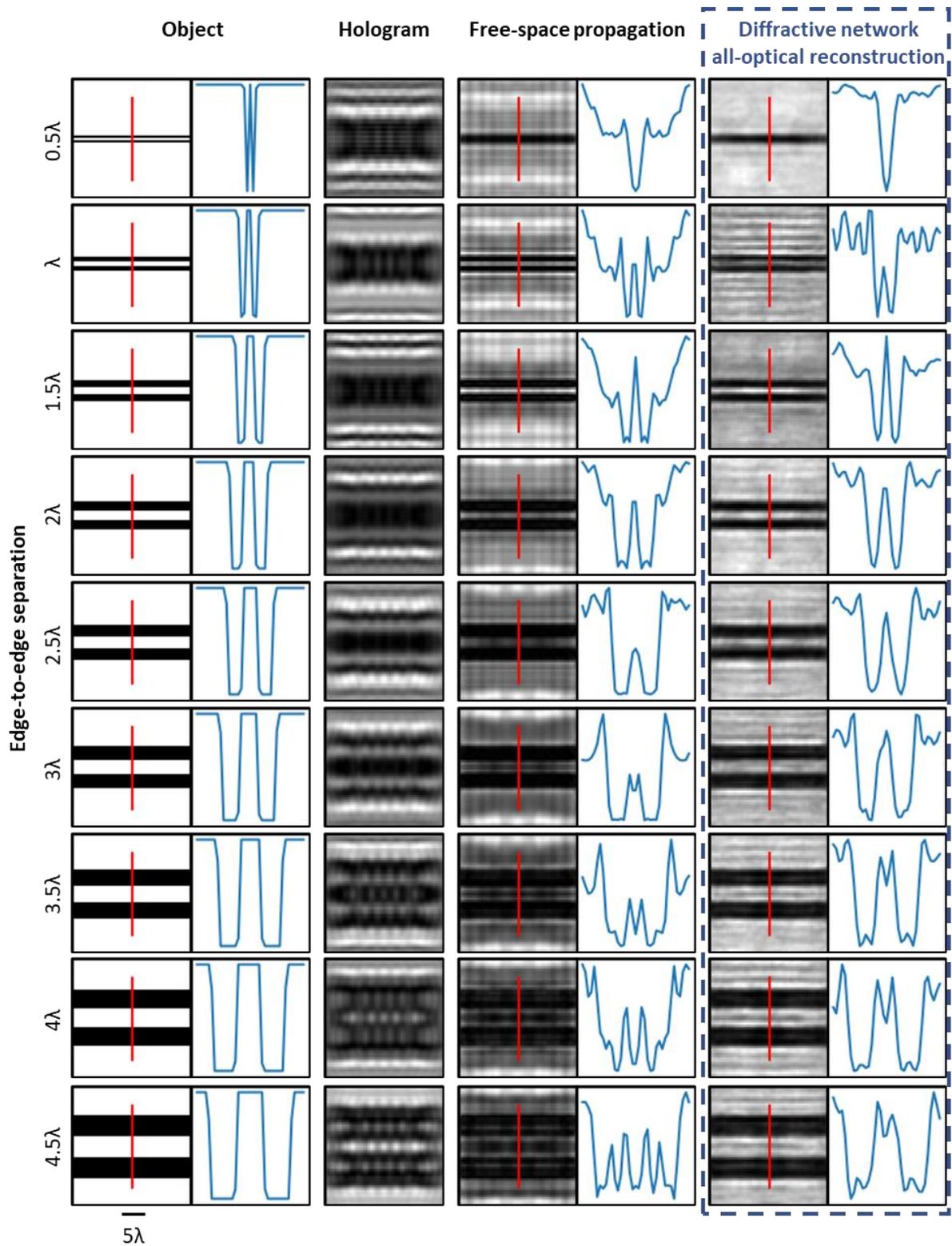

**Fig. 3 Quantification of all-optical holographic image reconstruction resolution.** The edge-to-edge separation refers to the separation from the inner edge of one line to the inner edge of the other. Intensity variations along the red lines are shown on the accompanying blue curves.

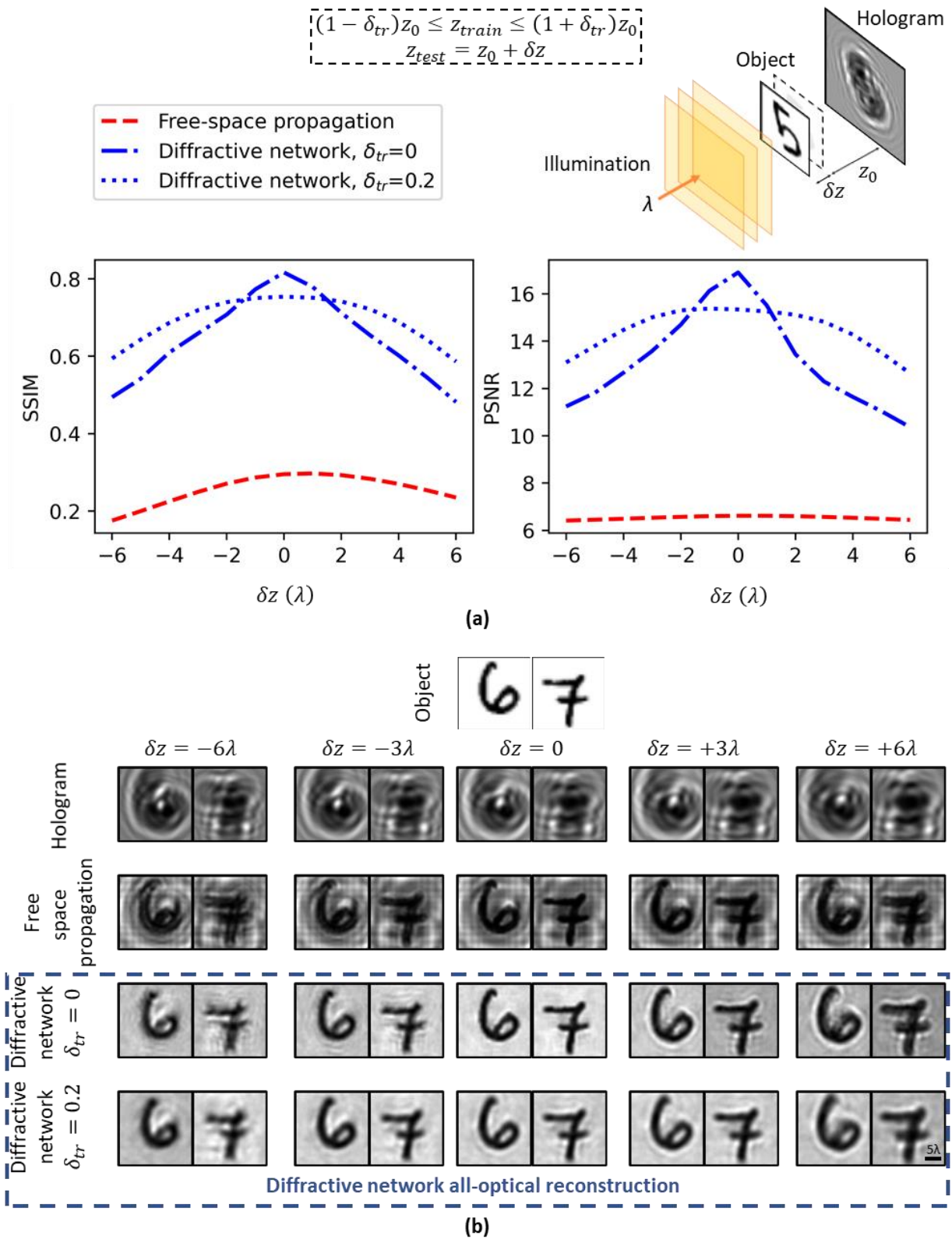

**Fig. 4 Robustness of all-optical diffractive network reconstructions against the hologram recording distance variations.** (a) Quantitative comparison of the robustness of the diffractive

network reconstructions and free-space propagation-based reconstructions to deviations in the hologram recording axial distance. SSIM and PSNR of the diffractive network results remain acceptable over a wide range of hologram recording distance variations. This robustness can be further improved by incorporating such random deviations in the training phase, as illustrated by the curves corresponding to $\delta_{tr} = 0.2$. These metrics (SSIM and PSNR) were computed over the 10,000 test images of MNIST dataset. (b) Visual comparison of the reconstructed images.

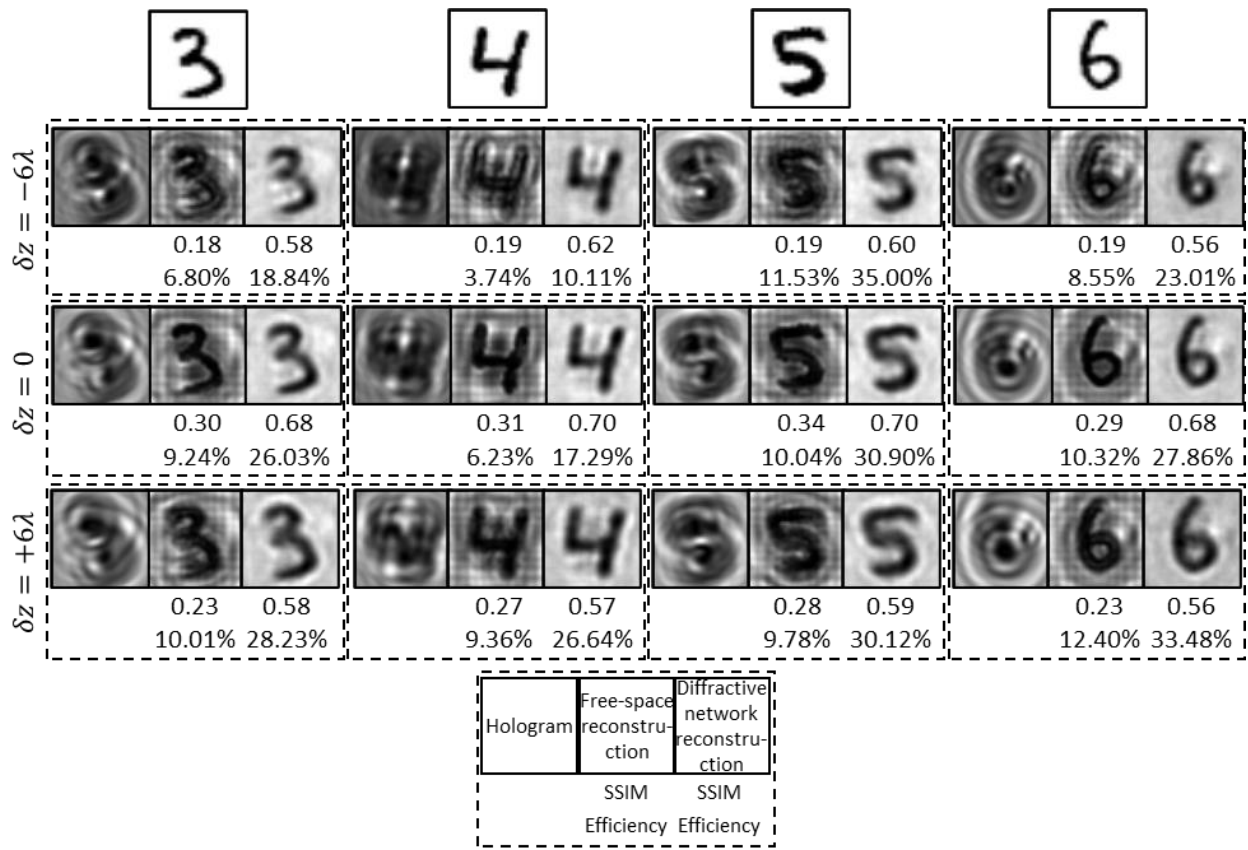

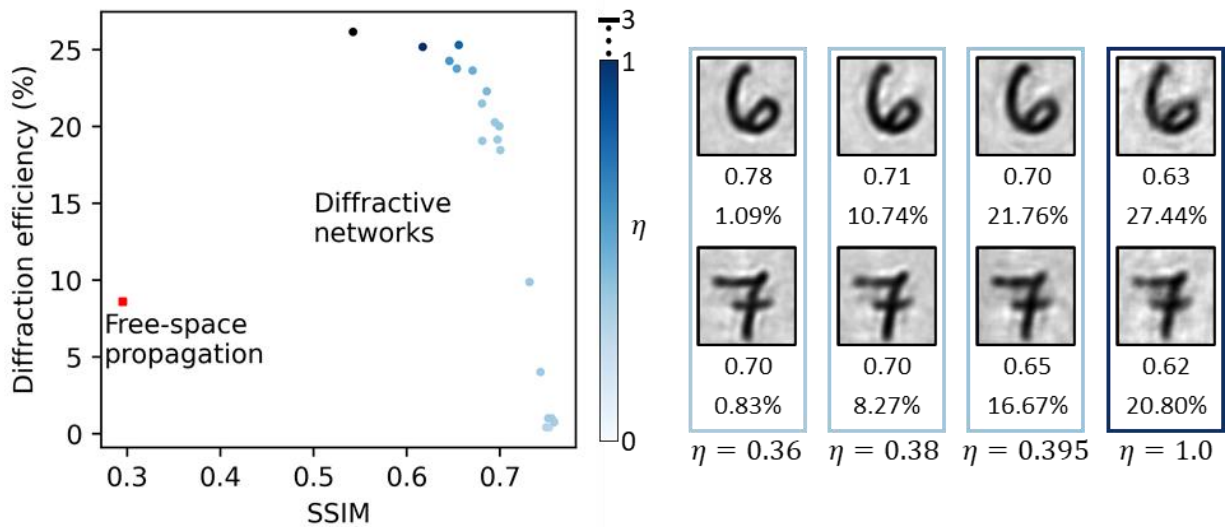

**Fig. 5 Diffraction efficiency improvement of all-optical holographic reconstructions performed by diffractive networks.** (**a**) Examples of all-optical holographic reconstructions by a high-efficiency diffractive network (with training hyperparameters: $\delta_{tr} = 0.2$, $\eta = 0.5$). For each panel in (a) corresponding to a recording $\delta z$, left: hologram, center: free-space propagation-based reconstruction,

right: diffractive network reconstruction. (**b**) SSIM vs. diffraction efficiency tradeoff achieved by tuning $\eta$ in the loss function. The two numbers below the reconstructed images correspond to the SSIM and the diffraction efficiency (%) of each reconstruction.